\documentclass[twocolumn,showpacs,preprintnumbers,amsmath,amssymb]{revtex4}
\usepackage{graphicx}% Include figure files
\usepackage{bm}% bold math

\newcommand{\im}{\mathrm{Im}\,}

\renewcommand{\d}{\mathsf{d}}
\newcommand{\opr}[1]{\mathsf{#1}}

\newcommand{\erfc}{\mathrm{erfc}}
\renewcommand{\[}{\begin{equation}}                               %
\renewcommand{\]}{\end{equation}}

\newcommand{\laplace}{\triangle}

\begin{document}

%\preprint{APS/123-QED}

\title{The decay law can have an irregular character}

\author{Pavel Exner$^{1,2}$ and Martin Fraas$^{1}$}
\address{
$^{1}$Nuclear Physics Institute, Czech Academy of Sciences,
25068 \v{R}e\v{z} near Prague, Czechia \\
$^{2}$Doppler Institute, Czech Technical University,
B\v{r}ehov\'{a} 7, 11519 Prague, Czechia }


\date{\today}

\begin{abstract}
Within a well-known decay model describing a particle confined
initially within a spherical $\delta$ potential shell, we consider
the situation when the undecayed state has an unusual energy
distribution decaying slowly as $k\to\infty$; the simplest example
corresponds to a wave function constant within the shell. We show
that the non-decay probability as a function of time behaves then
in a highly irregular, most likely fractal way.
\end{abstract}

\maketitle

The decay of an unstable quantum systems is one of the effects
frequently discussed and various aspects of such processes were
considered. To name just a few, recall the long-time deviation
from the exponential decay law\cite{longtime}, the short-time
behavior related to the Zeno and anti-Zeno effects \cite{BN, MS,
Sch, aZ}, revival effects such the classical one in the
kaon-antikaon system, etc. In all the existing literature
\cite{lit}, however, the decay law is treated as a smooth
function, either explicitly or implicitly, e.g. by dealing with
its derivatives. The aim of this letter is to show there are
situation when it is not the case.

A hint why it could be so comes from the behavior of Schr\"odinger
wave functions during the time evolution. While in most cases the
evolution causes smoothing \cite{smooth}, it may not be true for
for a particle confined in a potential well and the initial state
does not belong to the domain of the Hamiltonian. A simple and
striking example was found by M.~Berry \cite{MVB} for a
rectangular hard-wall box, and independently by B.~Thaller
\cite{THA} for a one-dimensional infinite potential well. It
appears that if the initial wave function is constant, it evolves
into a steplike-shaped $\psi(x,t)$ for times  which are rational
multiples of the period, $t=qT$ with $q=N/M$, and the number of
steps increases with growing $M$, while for an irrational $q$ the
function $\psi(x,t)$ is fractal w.r.t. the variable $x$.

One can naturally ask what will happen if the hard wall is
replaced by a semitransparent barrier through which the particle
can tunnel into the outside space. In a broad sense this is one of
the most classical decay model which can be traced back to
\cite{Gamow}. We will deal with its particular case when the
barrier is given by a spherical $\delta$ potential which is
sometimes called \emph{Winter model} being introduced for the
first time, to our knowledge, in \cite{winter}; see also
\cite{spheres}. The described behavior of the wave function in the
absence of tunneling suggests that in the decaying system the
irregular time dependence could also be visible, both in the wave
function and in various quantities derived from it \cite{VDS}, at
least in the weak coupling case. The aim of the present letter is
to demonstrate that this conjecture is indeed valid.

To be concrete, we will study a spinless nonrelativistic quantum
particle described by the Hamiltonian
 % ------------- %
 \begin{equation} \label{Ham}
 H_\alpha = -\laplace + \alpha \delta(|\vec{r}|-R)\,, \quad \alpha >0\,,
 \end{equation}
 % ------------- %
with a fixed $R>0$; we use rational units, $\hbar = 2m = 1$. For
simplicity we restrict our attention to the $s$-wave part of the
problem, writing thus the wave functions as $\psi(\vec{r},t)
=\frac{1}{\sqrt{4\pi}}r^{-1}\phi(r,\,t)$ with the associated
Hamiltonian
 % ------------- %
\[\opr{h}_\alpha = -\frac{\d^2}{\d r^2} + \alpha
\delta(r - R) \label{eq:parham}\]
 % ------------- %
in the lowest partial wave. We are interested in the time
evolution determined by the Hamiltonian (\ref{Ham}),
$\psi(\vec{r},t)=e^{-iH_\alpha t}\psi(\vec{r},0)$ for a fixed
initial condition $\psi(\vec{r},0)$ with the support inside the
ball of radius $R$; the advantage of the used model is that the
propagator can be computed explicitly. Of a particular interest is
the decay law,
 % ------------- %
 \begin{equation} \label{declaw}
 P(t) = \int\limits_0^R |\phi(r,\,t)|^2\,\d r\,,
 \end{equation}
 % ------------- %
i.e. the probability that the system localized initially within
the shell will be still found there at the measurement performed
at an instant $t$. We are going to derive an exact formula for the
decay law which we will then allow us to evaluate the function
(\ref{declaw}) numerically for a given initial state.

It is straightforward to check \cite{spheres} that the Hamiltonian
(\ref{Ham}) has no bound states. On the other hand, it has
infinitely many resonances with the widths increasing
logarithmically w.r.t. the resonance index. A natural and well
known idea \cite{GMMMMG, GMM} is to employ them as a tool to
expand the quantities of interest.

First of all, we have to find Green's function $g(k,\,r,\,r')$,
i.e. the integral kernel of $(\opr{h}_\alpha -k^2)^{-1}$ which
determines the time evolution in the standard way \cite{RS},
 % ------------- %
 \begin{equation} \label{eq:timeop}
 e^{-i\opr{h}_\alpha t} = \frac{1}{\pi}
 \lim_{\varepsilon \downarrow 0} \int\limits_0^\infty
 e^{-i \lambda t}\:
 \im \frac{1} {\opr{h}_\alpha - \lambda - i \varepsilon}\:
 \d \lambda\,,
 \end{equation}
 % ------------- %
recall that $\sigma(\opr{h}_\alpha)= [0,\infty)$ for $\alpha>0$;
we are going to perform a resonance expansion of the integral
(\ref{eq:timeop}). The Green function for a system with singular
potential is obtained from Krein's formula,
 % ------------- %
 $$
 \frac{1}{\opr{h}_\alpha - k^2} = \frac{1}{\opr{h}_0 - k^2} +
 \lambda(k) (\Phi_k,\cdot)\Phi_k(r)\,, \label{eq:kreinfor}
 $$
 % ------------- %
where $\Phi_k(r)$ is the free Green function, in particular,
$\Phi_k(r) = \frac{1}{k}\, \sin(kr)\, e^{ikR}$ holds for $r<R$,
and $\lambda(k)$ is determined by boundary conditions at the
singular point $R$; by a direct calculation \cite{spheres} one
finds
 % ------------- %
 \begin{equation} \label{eq:resolvent}
\lambda(k) = -\frac{\alpha}{1 + \frac{i \alpha}{2 k}(1 -
e^{2ikR})}\,.
 \end{equation}
 % ------------- %
This allows us to write the sought reduced wave function at time
$t$ as $\phi(r,\,t) = \int\limits_0^\infty u(t,\,r,\,r')
\phi(r',\,0) \,\d r'$ with
 % ------------- %
 $$
u(t,\,r,\,r') =  \frac{1}{\pi} \lim_{\varepsilon \downarrow 0}
\int\limits_0^\infty e^{-ik^2t}\: \im g(k+ i \varepsilon,\,r,\,r')
\: 2k\, \d k
 $$
 % ------------- %
Since $\opr{h}_\alpha$ has no eigenvalues we may pass in the last
formula to the limit $\varepsilon \to 0 $ obtaining
 % ------------- %
\[ u(t,\,r,\,r') = \int\limits_0^\infty p
(k,\,r,\,r') \exp(-ik^2t)\: 2 k\, \d k, \label{eq:timeprop2} \]
 % ------------- %
where $p(k,\,r,\,r') = \frac{1}{\pi} \im g(k,\,r,\,r')$. This can
be using equation (\ref{eq:resolvent}) written explicitly as
 % ------------- %
\[ p(k,\,r,\,r') = \frac{2k\sin(kr)\sin(kr')}{\pi(2 k^2 +
2\alpha^2 \sin^2 kR + 2 k \alpha \sin 2kR)}\,.
\label{eq:projekce}\]
 % ------------- %
The resonances of the problem are identified with the poles of
$\,g(\cdot,\,r,\,r')\,$ continued analytically to the lower
momentum halfplane. They exist in pairs, those in the fourth
quadrant, denoted as $k_n$ in the increasing order of their real
parts, and $-\bar{k}_n$. The set of singularities of the kernel
(\ref{eq:projekce}) then include also the mirrored points, being
 % ------------- %
\[ S = \{k_n,\,-k_n,\,\bar{k}_n,\,-\bar{k}_n :\:
n\in\mathbb{N}\}.\label{eq:9}\]
 % ------------- %
In the vicinity of the singular point $k_n$ the function
$p(\cdot,\,r,\,r')$ can be written as
 % ------------- %
\[p(k,\,r,\,r') = \frac{i}{2 \pi} \frac{v_n(r) v_n(r')}{k^2 -
k_n^2} + \chi(k,\,r,\,r'), \label{eq:10} \]
 % ------------- %
where $v_n(r)$ solves the differential equation $\opr{h}_\alpha
v_n(r) = k_n^2 v_n(r) $ and the function $\chi$ is locally
analytic.

The factor $\frac{i}{2 \pi}$ is chosen to get the conventional
normalization of the resonant state $v_n(r)\:$ \cite{GMMMMG}. The
last named paper also demonstrates that for a finite-range
potential barrier and $r,\,r' < R$ the function
$p(\cdot,\,r,\,r')$ decreases in every direction of the $k$-plane;
in the present case it is not difficult to verify this claim
directly. Then one can express $p(k,\,r,\,r')$ as the sum over the
pole singularities
 % ------------- %
\[ p(k,\,r,\,r') = \sum_{\tilde{k} \in S}\,
\frac{1}{k-\tilde{k}}\:
\mathrm{Res}_{\tilde{k}}\, p(k,\,r,\,r') \label{eq:11} \]
 % ------------- %
and derive from the residue theorem the following useful formula
 % ------------- %
\[ \sum_{\tilde{k} \in S}\, \mathrm{Res}_{\tilde{k}}\,
p(k,\,r,\,r') = 0\,. \label{eq:2.1} \]
 % ------------- %
The last two equations can be rewritten in view of eq.
(\ref{eq:10}) and the symmetry of the set $S$ in the form
 % ------------- %
\begin{eqnarray}
p(k,\,r,\,r') = \sum_{n \in \mathbb{Z}}\, \frac{i}{2 \pi}\,
\frac{1}{k^2-k^2_n}\, \frac{k}{k_n}\, v_n(r)
v_n(r')  \label{eq:12}\,, \\
\sum_{n \in \mathbb{Z}}\, \frac{1}{k_n}\, v_n(r)v_n(r') = 0\,,
\label{eq:2.2}
\end{eqnarray}
 % ------------- %
where we denote $k_{-n} := -\bar{k}_{n}$ and $v_{-n}$ is the
associated solution of the equation $\opr{H}_\alpha v_{-n}(r) =
k_{-n}^2 v_{-n}(r)$.

Next we substitute from eq. (\ref{eq:12}) into
(\ref{eq:timeprop2}) and using the identity (\ref{eq:2.2}) we
arrive at the formula
 % ------------- %
\begin{eqnarray}u(t,\,r,\,r') &=& \frac{i}{2 \pi}\int\limits_0^\infty
\sum_{n \in \mathbb{Z}} \frac{\exp(-ik^2t)}{k^2-k^2_n}
\frac{2k^2}{k_n} v_n(r)v_n(r')\d k \nonumber \\
&=& \sum_{n \in \mathbb{Z}}\, M(k_n,\,t) v_n(r) v_n(r')
\label{eq:13}
\end{eqnarray}
 % ------------- %
with $M(k_n,\,t) = \frac{1}{2}\, e^{u_n^2}\, \erfc(u_n)$ and
$u_n:= -e^{-i\pi/4} k_n \sqrt{t}$.
%\[M(k_n,\,t) = \frac{1}{2}\, e^{u_n^2}\, \erfc(u_n)\,,\quad
%u_n:= -e^{-i\pi/4} k_n \sqrt{t}\,. \label{eq:14}\]
 % ------------- %
Indeed, using $2k^2= 2(k^2-k^2_n) +2k_n^2$ we write
$u(t,\,r,\,r')$ as a sum of two terms the first of which vanishes
in view of (\ref{eq:2.2}). The second one decomposes again into a
sum of two integrals containing $k_n\pm k$ in their denominators,
which gives the right-hand side of (\ref{eq:13}) with
 % ------------- %
\[M(k_n,\,t) = \frac{i}{2 \pi} \int\limits_{-\infty}^\infty
\frac{e^{-ik^2t}}{k-k_n}\, \d k\, \label{eq:15},\]
 % ------------- %
in other words, the above expression.

Now a straightforward calculation using (\ref{declaw}) and
(\ref{eq:13}) allows us to express the decay law in the form
 % ------------- %
\[ P(t) = \sum_{n,\,l} C_n \bar{C}_l I_{nl} M(k_n,\,t)
\overline{M(k_l,\,t)} \label{eq:16}\,, \]
 % ------------- %
with the coefficients
 % ------------- %
\begin{equation}
  C_n := \int\limits_0^R \phi(r,\,0) v_n(r)\, \d r \,, \;
  I_{nl} := \int\limits_0^R v_n(r) \bar{v}_l(r)\, \d r\,. \label{eq:17}
\end{equation}
 % ------------- %
This expression holds generally, see \cite{GMM}; in our particular
case of the Hamiltonian (\ref{eq:parham}) we can specify $v_n(r) =
\sqrt{2} Q_n\sin(k_n r)$ with the coefficient $Q_n$ equal to
 % ------------- %
 $$ \left(\frac{-2i k_n^2}{2k_n + \alpha^2 R \sin 2 k_nR + \alpha\sin 2k_nR
+ 2 k_n \alpha R \cos 2k_nR}\right)^{1/2}
 $$
 % ------------- %

Now we are ready to compute the decay law for a given initial
state. Without loss of generality we may put $R=1$, we choose the
value $\alpha = 500$ for numerical evaluation and replace the
infinite series by a cut-off one with $|n|\le 1000$. As the first
example we consider initial wave function constant within the
well, i.e.
 % ------------- %
\[
  \phi(r,\,0) = R^{-3/2}\sqrt{3} r\,,\quad r<R\,.
 \label{eq:4.1}
\]
 % ------------- %
\begin{figure}
\includegraphics[width=\linewidth,keepaspectratio]{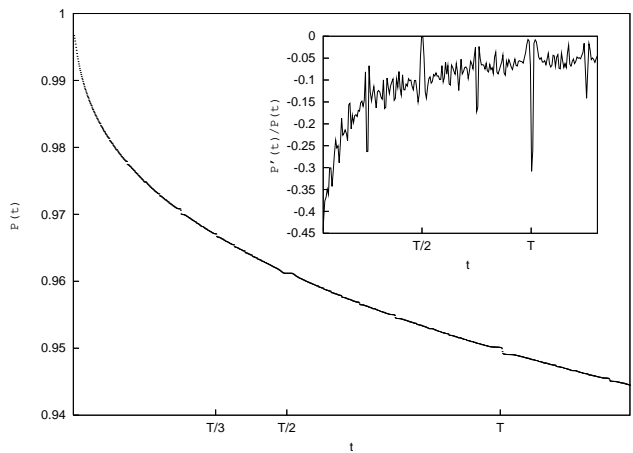}
\caption{\label{fig:1} Decay law for the initial state
$\phi(r,\,0) = R^{-3/2}\sqrt{3} r$. In the inset we plot the
logarithmic derivative averaged over intervals of the length
approximately $T/200$.}
\end{figure}
 % ------------- %
The corresponding decay law is plotted in Figure~\ref{fig:1}. It
is irregular having numerous steps \cite{antizeno}, the most
pronounced at the period $T=2R^2/\pi$ and its simple rational
multiples. To make them more visible we plot in the inset the
logarithmic derivative of the function $P(t)$; it is locally
smeared, otherwise the picture would be a fuzzy band. The
irregular structure is expected to be fractal; it persists at
higher time but its amplitude decreases relatively w.r.t. the
smooth background.

\begin{figure}
\includegraphics[width=\linewidth,keepaspectratio]{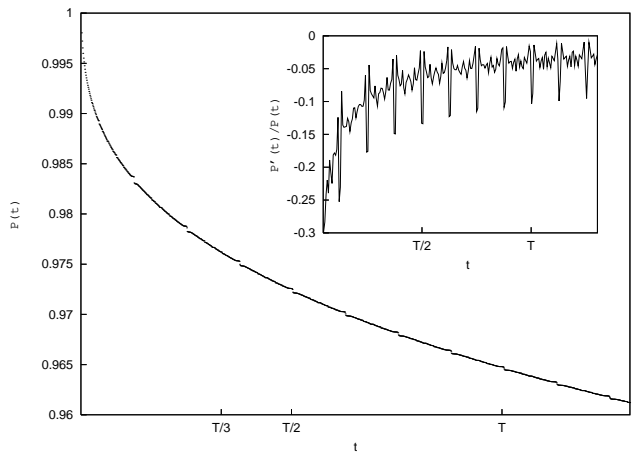}
\caption{\label{fig:2} Decay law for initial state $\phi(r,\,0) =
R^{-1/2}$ and its logarithmic derivative, locally smeared, in the
inset.}
\end{figure}

\begin{figure}
\includegraphics[width=\linewidth,keepaspectratio]{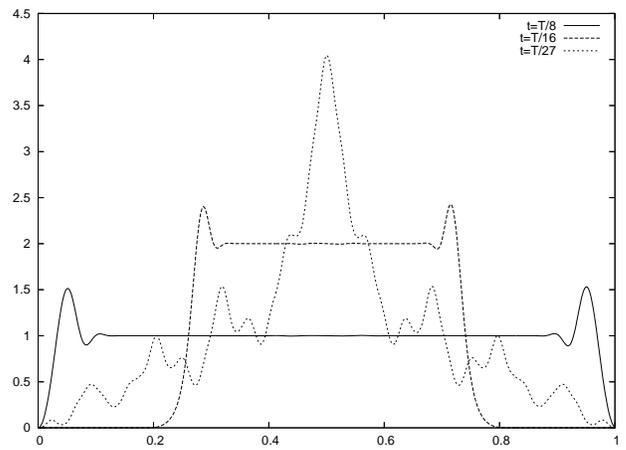}
\caption{\label{fig:3} The probability density inside the sphere
with $R=1$ multiplied by $r^2$ at the instants $t=T/8,\,T/16$, and
$T/27$.}
\end{figure}

In the next example we choose the initial state having constant
reduced wave function, $\phi(r,\,0) = R^{-1/2}$ for $r<R$, so
$\psi(r,\,0)$ has a (square integrable) singularity at the origin;
the advantage is that the reduced problem offers a straightforward
comparison to the one-dimensional example treated in \cite{THA}
including the shapes of the wave functions. The decay law for this
case is plotted in Figure~\ref{fig:2}; it again exhibits
derivative jumps around simple rational multiples of the period.
 % ------------- %
 % ------------- %
The corresponding function $|\phi(r,\,t)|^2$ for three such values
is plotted in Figure~\ref{fig:3}.
 % ------------- %
 % ------------- %
To compare with \cite{THA} notice that the one-dimensional well
has length $R=1$ so the revival period of this state is $T/8$. In
the absence of decay the function for $t=T/8$ is just constant,
the other two are simple step functions. We see that the tunneling
through $\delta$-barrier modifies the shape of the function mostly
in the vicinity of the origin, the barrier, and the jump points.

To support the conjecture about the fractal character of $P(t)$
let us look how its derivative behaves in the limit $\alpha \to
\infty$ when $\phi(r,\,t)$ expands asymptotically as
 % ------------- %
 \begin{equation}
 \phi(r,\,t) \approx
\sum_{n} C_n \exp(-i k_n^2 t) v_n(r) \label{eq:waveasympt}\,.
 \end{equation}
 % ------------- %
It is easy to see that for a fixed $n$ and $\alpha \to \infty$ the
resonance position expands around $k_{n,0}:=n\pi/R$ as
 % ------------- %
\begin{equation}
  k_n \approx\, k_{n,0} - \frac{k_{n,0}}{\alpha R}
  + \frac{k_{n,0}}{(\alpha R)^2} -i\frac{k^2_{n,0}}{\alpha^2 R}
  + \cdots \label{eq:resexp}
\end{equation}
 % ------------- %
which shows that in the leading order we have $v_n(r) \approx
\sqrt{\frac{2}{R}} \sin(k_n r)$, and furthermore, that the
substantial contribution to the sum in (\ref{eq:waveasympt}) comes
from terms with $ n \lesssim
[\alpha^{1-\varepsilon}\frac{R}{\pi}]$ for some
$0<\varepsilon<1/3$.

The derivative $\dot{P}(t)$ can be computed from the probability
current conservation $j(r)= -\frac{\d}{\d t} |\phi(r,\,t)|^2$;
integrating it over the interval $(0,R)$ and using $j(0) = 0$ we
get
 % ------------- %
 \[ \dot{P}(t) = -2 \im (\phi'(R,\,t) \bar{\phi}(R,\,t))\,. \label{eq:pder} \]
 % ------------- %
We plug the above expansion into (\ref{eq:waveasympt}) obtaining
 % ------------- %
\begin{eqnarray}
 \phi(R,\,t) &\approx& \sqrt{\frac{2}{R}} \sum_{n=1}^{\infty} (-1)^n C_n
 \exp\left[-i k^2_{n,0} t\left(1-\frac{2}{\alpha R} \right)\right] \nonumber \\ &&
 \exp\left(-\frac{2k^3_{n,0}}{\alpha^2 R}t\right)
 \left(-\frac{k_{n,0}}{\alpha} - i\frac{k^2_{n,0}}{\alpha^2} \right)
 \label{eq:asymstate}
\end{eqnarray}
 % ------------- %
and a similar expansion for $\phi'(R,\,t)$ with the last bracket
replaced by $k_{n,0}$. We observe that for $j>-1$ we have
 % ------------- %
 $$ \sum_{n=1}^{\infty} \exp\left(-2\frac{k^3_{n,0}}{\alpha^2 R}t \right)
k^j_{n,0} \approx \frac{R}{\pi}
\left(\frac{R}{2t}\right)^{(j+1)/3} \alpha^{2(j+1)/3} I_j\,,
 $$
 % ------------- %
where we have denoted $I_j := \int\limits_0^\infty e^{-x^3}\,x^j\,
\d x = \frac{1}{3}\Gamma\left(\frac{j+1}{3} \right)$.

Let us now we assume that the coefficients in
(\ref{eq:waveasympt}) satisfy $C_n \sim k^{-p}_{n,0}$ as
$n\to\infty$. Suppose first that the decay is fast enough, $p>1$;
notice that this is certainly true in the finite-energy case with
$p>3/2$. The term $-k_{n,0}/\alpha$ in (\ref{eq:asymstate})
obviously does not contribute to the imaginary part, hence we find
that $|\dot{P}(t)| \leq \mathrm{const}\: \alpha^{4/3 - 4/3p} \to
0$ holds as $\alpha\to\infty$ uniformly in the time variable.

The situation is different if the decay is slow, $p \leq 1$. As an
illustration take $ C_n = (-1)^{n+1} \frac{\sqrt{6}}{R k_n}$,
which corresponds to the first one of the above numerical
examples. Since the real part of the resonances changes with
$\alpha$, cf.~(\ref{eq:resexp}), it is natural to study the limit
of $\dot{P}(t_\alpha)$ as $\alpha\to\infty$ at the moving time
value $t_\alpha := t (1+2/\alpha R)$. Up to higher order terms the
appropriate value $\phi(R,\,t_\alpha)$ is obtained by removing the
bracket $(1-2/\alpha R)$ at the right-hand side of
(\ref{eq:asymstate}) and $\phi'(R,\,t_\alpha)$ is obtained
similarly.

Consider first irrational multiples of $T$. We use the observation
made in \cite{BG} that the modulus of $\sum_{n=1}^{L} e^{i\pi n^2
t}$ is for an irrational $t$ bound by $C\, L^{1-\varepsilon}$
where $C, \varepsilon$ depend on $t$ only. In combination with a
Cauchy-like estimate, $\sum_{n=1}^\infty a_n b_n \leq
\sum_{n=1}^\infty |\sum_{j=1}^n a_j|\, |b_{n} - b_{n+1}|$ which
yields
 % ------------- %
 $$
 \sum_{n=1}^\infty
 \exp(-ik^2_{n,0}t) \exp\left(-\frac{2k^2_{n,0}}
 {\alpha^2 R}t \right)
 k^j_{n,0} \lesssim \mathrm{const}\,
 \alpha^{2/3(j+1-\varepsilon)}\,
 $$
 % ------------- %
and consequently, $\dot{P}(t_\alpha)\to 0$ as $\alpha\to\infty$
similarly as in the case of fast decaying coefficients.

Let us assume next rational times, $t = \frac{p}{q}\, T$. If $pq$
is odd then $S_L(t):=\sum_{n=1}^{L} e^{i\pi n^2 t}$ repeatedly
retraces by \cite{BG} the same pattern, hence $\dot{P}(t_\alpha)
\to 0$ -- cf. Fig.~\ref{fig:1} at the half period. On the other
hand, for $pq$ even $|S_L(t)|$ grows linearly with $L$, and
consequently, $\lim\limits_{\alpha \to \infty} \dot{P}(t_\alpha)
>0$. As a example let us compute this limit for the period
$T$, i.e. $p=q=1$. Using (\ref{eq:pder}) we find
%-------------------
\begin{multline}
  \lim_{\alpha \to \infty}\dot{P}(T_\alpha) =
-\frac{24}{R^2} \lim_{\alpha \to \infty}
\im\left(\sum_{n=1}^{\infty} \exp\left(-\frac{2k^3_{n,0}}{\alpha^2
  R}T\right) \right. \\
\left. \sum_{n=1}^\infty \exp\left(-\frac{2k^3_{n,0}}{\alpha^2
  R}T\right)\left(-\frac{1}{\alpha} + i \frac{k_{n,0}}{\alpha^2}
  \right) \right) \\
  = -\frac{24}{R^2}\left(\frac{R}{\pi}\right)^2 \frac{1}{2T/R}\, I_1 I_0 =
  -\frac{4}{3 \sqrt{3}} \approx -0.77\,; \nonumber%\label{eq:5.332}
\end{multline}
%------------------
notice that without a local smearing $\dot P(T)$ in the inset of
Fig.~\ref{fig:1} would take approximately this value.

In conclusion, we have reexamined time decay in Winter model and
found indications that the decay law is a highly irregular
function if the energy distribution decays slowly as $k\to\infty$.

\medskip The research was partially supported by GAAS and MEYS of
the Czech Republic under projects A100480501 and LC06002.

%\bibliography{apssamp}% Produces the bibliography via BibTeX.

\begin{thebibliography}{99}

\bibitem{longtime}
L.A.~Khalfin, \emph{Zh. Eksp. Teor. Fiz.} \textbf{33}, 1371
(1957); for a more complete bibliography see Ref.~\cite{open}

\bibitem{open} P.~Exner: \emph{Open Quantum Systems and Feynman
Integrals}, D.~Reidel, Dordrecht 1985.

\bibitem{BN}
J.~Beskow, J.~Nilsson, {\em Arkiv Fys.} {\bf 34}, 561 (1967).

\bibitem{MS}
B.~Misra, E.C.G.~Sudarshan, \emph{J. Math. Phys.} {\bf 18}, 756
(1977).

\bibitem{Sch}
A.U.~Schmidt, in {\em Mathematical Physics Research on Leading
Edge} (Ch.~Benton, ed.), Nova Sci, Hauppauge NY, 2004;
pp.~113-143.

\bibitem{aZ}
P.~Exner, \emph{J. Phys.} {\bf A38}, L449 (2005).

\bibitem{lit}
The bibliography concerning unstable systems is vast. A large part
can be derived from sources like Refs.~2 and 5; one can also
mention, e.g., L.~Fonda, G.C.~Ghirardi, A.~Rimini, \emph{Rep.
Progr. Phys.} \textbf{41}, 587 (1978).

\bibitem{smooth}
K.~Yajima, in \emph{Functional-analytic methods for partial
differential equations},  Springer Lecture Notes in Math.,
vol.~1450, Berlin 1990; p.~20

\bibitem{MVB}
M.V.~Berry, \emph{J. Phys.} \textbf{A29}, 6617 (1996).

\bibitem{THA} B.~Thaller: \emph{Visual Quantum Mechanics},
Springer, Heidelberg 2000.

\bibitem{Gamow}
G.~Gamow, \emph{Z. Phys.} \textbf{51}, 204 (1928).

\bibitem{winter}
R.G.~Winter, \emph{Phys. Rev.} \textbf{123}, 1503 (1961).

\bibitem{spheres}
A thorough analysis of Winter's model can be found in
J.-P.~Antoine, F.~Gesztesy, J.~Shabani, \emph{J. Phys.}
\textbf{A20}, 3687 (1987). There are various generalizations of
the model; we refer to the S.~Albeverio, F.~Gesztesy, R.~H\o
egh-Krohn, H.~Holden: \emph{Solvable Models in Quantum mechanics},
2nd edition, AMS Chelsea 2005, for a bibliography.

\bibitem{VDS}
This conjecture is also supported by the study of revivals in an
infinite square well with a $\delta$ barrier, see G.A.~Vugalter,
A.K.~Das, V.A.~Sorokin, \emph{Phys. Rev.} \textbf{A66}, 012104
(2002).

\bibitem{GMMMMG}
G.~Garc\'{\i}a-Calder\'on, R.~Peierls, \emph{Nucl. Phys.}
\textbf{A265}, 443 (1976).

\bibitem{GMM}
G.~Garc\'{\i}a-Calder\'on, J.L.~Mateos, M~.Moshinsky, \emph{Phys.
Rev. Lett.} \textbf{74}, 337 (1995).

\bibitem{RS}
M.~Reed, B.~Simon: \emph{Methods of Modern Mathematical Physics I.
Functional Analysis}, Academic Press, New York 1972.

\bibitem{antizeno}
Another striking deviation from the exponential decay law is that
the decay rate explodes as $t\to 0$, which is due to the fact that
the energy distribution of the state $\psi$ decays too slowly at
high energies -- cf.~P.~Exner, J. Phys. \textbf{A38}, L449 (2005).

\bibitem{BG}
M.~V.~Berry, J.~Goldberg, \emph{Nonlinearity} \textbf{1}, 1
(1988).

\end{thebibliography}

\end{document}